\documentclass{aamas2016}

\usepackage{algorithm, algorithmic}

\begin{document}


\title{Resource Abstraction for Reinforcement Learning\\in Multiagent Congestion Problems\\}



%
%
%
%

%

\numberofauthors{3}

\author{
%
\alignauthor
Kleanthis Malialis\titlenote{Most of this work was carried out when the author was at the University of York, UK.}\\
       \affaddr{Dept. of Computer Science\\University College London, UK}\\
       \email{k.malialis@ucl.ac.uk}
\alignauthor
Sam Devlin\\
       \affaddr{Dept. of Computer Science\\University of York, UK}\\
       \email{sam.devlin@york.ac.uk}
\alignauthor
Daniel Kudenko\\
       \affaddr{Dept. of Computer Science\\University of York, UK}\\
       \email{daniel.kudenko@york.ac.uk}
}

\maketitle

\begin{abstract}
Real-world congestion problems (e.g. traffic congestion) are typically very complex and large-scale. Multiagent reinforcement learning (MARL) is a promising candidate for dealing with this emerging complexity by providing an autonomous and distributed solution to these problems. However, there are three limiting factors that affect the deployability of MARL approaches to congestion problems. These are learning time, scalability and decentralised coordination i.e. no communication between the learning agents. In this paper we introduce Resource Abstraction, an approach that addresses these challenges by allocating the available resources into abstract groups. This abstraction creates new reward functions that provide a more informative signal to the learning agents and aid the coordination amongst them. Experimental work is conducted on two benchmark domains from the literature, an abstract congestion problem and a realistic traffic congestion problem. The current state-of-the-art for solving multiagent congestion problems is a form of reward shaping called difference rewards. We show that the system using Resource Abstraction significantly improves the learning speed and scalability, and achieves the highest possible or near-highest joint performance/social welfare for both congestion problems in large-scale scenarios involving up to 1000 reinforcement learning agents.
\end{abstract}


%
%

%
%






\section{Introduction}
There has been in recent years an explosion of interest in multiagent systems (MAS) where multiple interacting agents are situated in a common environment. A larger set of problem domains can be practically modelled using MAS for example by taking advantage of the geographic distribution and sharing experiences for faster and better learning \cite{stone2000}. MAS often need to be very complex, and MARL is a promising candidate for dealing with this emerging complexity \cite{stone2007}. MARL approaches provide adaptive and autonomous agents that can improve from experience.

Congestion problems have been formulated in the past as MARL problems \cite{tumer2007,tumer2009,proper2013,devlin2014}. Congestion problems are typically very complex and large-scale. Real-life examples include traffic congestion, air traffic management and network routing. Despite the benefits offered by MARL approaches there are three limiting factors that affect their deployability to real-world congestion problems; these are, learning time, scalability and decentralised coordination i.e. no communication between learning agents.

Firstly, learning time is of particular importance to offline learning and critical to online learning. If a MARL system is slow it may be practically useless even if it eventually achieves a great performance. Furthermore, in real-life complex congestion problems, learning time is typically limited.

Secondly, the curse of dimensionality \cite{sutton1998} refers to the exponential growth of the search space, in terms of the number of states and actions. If a MARL solution is not scalable, it will never be considered, let alone adopted, for deployment by a company or organisation.

Lastly, while sharing experiences may be beneficial and even necessary in specific applications, allowing communication between the learning agents is not always desirable or even possible. This is because agents may have been developed by different designers or controlled by different owners. Even if this is not the case, communication messages can be costly, noisy, corrupted, dropped or received out of order \cite{mataric1998}.

In this paper we introduce Resource Abstraction, an approach that addresses the aforementioned challenges.  Experiments are conducted on two benchmark domains from the literature; an abstract congestion problem and a realistic traffic congestion problem where agents or cars ``compete'' for traffic lanes. Our contributions are the following.

We propose the novel Resource Abstraction approach where its main idea is to allocate the available resources into groups called abstract groups. This grouping provides an abstract ($A$) reward function that is more informative and aids the coordination among the learning agents to significantly improve the learning speed, scalability and agents' final joint performance or social welfare.\\

We compare our approach against systems which learn from local ($L$), global ($G$) and difference ($D$) rewards. The first two are the common approaches while the third is a form of reward shaping that constitutes the state-of-the-art for solving multiagent congestion problems.

Like $D$, the system using $A$ provides decentralised coordination. It is further shown that the system using $A$ requires significantly fewer time steps per learning episode to achieve the highest system performance or social welfare in large-scale scenarios. The proposed approach scales to a large number of agents and is successfully demonstrated in experiments involving up to 1000 learning agents.

The organisation of this paper is as follows. Section~\ref{sec:background} describes the background material necessary to understand the contributions made by this work. Section~\ref{sec:formulation} presents a formal definition of congestion problems as multiagent learning and coordination problems. Section~\ref{sec:abstract} introduces in detail our proposed Resource Abstraction approach. Section~\ref{sec:BPD} and Section~\ref{sec:TLD} provide the experimental work and results for the abstract and traffic congestion problems respectively. A discussion and conclusion is presented in Section~\ref{sec:conclusion}.

\section{Background}\label{sec:background}

\subsection{Reinforcement Learning}
Reinforcement learning is a paradigm in which an active decision-making agent interacts with its environment and learns from reinforcement, that is, a numeric feedback in the form of reward or punishment \cite{sutton1998}. The feedback received is used to improve the agent's actions. The problem of solving a reinforcement learning task is to find a policy (i.e. a mapping from states to actions) which maximises the accumulated reward.

The concept of an iterative approach constitutes the backbone of the majority of reinforcement learning algorithms. These algorithms apply so called temporal-difference updates to propagate information about values of states, $V(s)$, or state-action, $Q(s,a)$, pairs. These updates are based on the difference of the two temporally different estimates of a particular state or state-action value. The Q-learning algorithm is such a method \cite{watkins1992}. After each real transition, $(s,a) \rightarrow (s^{\prime},r)$, in the environment, it updates state-action values by the formula:
\begin{equation}\label{eq:qlearning}
  Q(s,a)\leftarrow Q(s,a)+\alpha[r+\gamma \max_{a'} Q(s',a')-Q(s,a)]
\end{equation}
where $\alpha$ is the rate of learning and $\gamma$ is the discount factor.

The exploration-exploitation trade-off constitutes a critical issue in the design of a reinforcement learning agent. It aims to offer a balance between the exploitation of the agent's knowledge and the exploration through which the agent's knowledge is enriched. A common method of doing so is $\epsilon$-greedy, where the agent behaves greedily most of the time, but with a probability $\epsilon$ it selects an action randomly. To get the best of both exploration and exploitation, it is advised to reduce $\epsilon$ over time \cite{sutton1998}.

Applications of MARL typically take one of two approaches; multiple individual learners (ILs) or joint action learners (JALs) \cite{claus1998}. Multiple ILs assume any other agents to be part of the environment and so, as the others simultaneously learn, the environment appears to be dynamic as the probability of transition when taking action $a$ in state $s$ changes over time. To overcome the appearance of a dynamic environment, JALs were developed that extend their value function to consider for each state the value of each possible combination of actions by all agents. The consideration of the joint action causes an exponential increase in the number of values that must be calculated with each additional agent added to the system. Therefore, as we are interested in scalability and decentralised coordination (i.e. no communication between agents), this work focuses on multiple individual learners and not joint action learners.

In MARL the typical approach provides agents with one of two types of reward. A \textbf{local reward ($L$)} is unique to each agent and often promotes ``selfish'' behaviours since each agent attempts to increase its own reward, potentially at the cost of the system performance. The \textbf{global reward ($G$)} is in fact the system performance used as a learning signal\footnote{The terms global reward, system performance and social welfare are used interchangeably.}. This allows each agent to act in the system's interest and typically, the learnt behaviour is better compared to the case where agents are learning using $L$.

$G$, however, includes a substantial amount of noise due to other agents acting in the system. This is because an agent may be rewarded for taking a bad action (if the other agents executed a good action), or punished for taking a good action (if the others performed a bad action). This is known as the multiagent credit assignment problem. The reward signal should reward an agent depending on its individual contribution to the system objective. Difference rewards \cite{wolpert2000}, a form of reward shaping, address this issue and are discussed in the next section.

\subsection{Difference Rewards}\label{sec:D}
Difference rewards \cite{wolpert2000} were introduced to address the multiagent credit assignment problem encountered in reinforcement learning. The \textbf{difference rewards ($D_i$)} is a shaped reward signal that helps an agent $i$ to learn the consequences of its actions on the system objective by removing a large amount of the noise created by the actions of other agents active in the system. It is defined as:
\begin{equation}
D_{i}(z) = G(z) - G(z_{-i})
\end{equation}
where $z$ is a general term representative of either states or state-action pairs depending on the application, $G(z)$ is the global reward used, and $G(z_{-i})$ is $G(z)$ for a theoretical system without the contribution of agent $i$.

Difference rewards exhibit the following two properties. Firstly, any action taken that increases $D_i$ simultaneously increases $G$. Secondly, since difference rewards only depend on the actions of agent $i$, $D_i$ provides a cleaner signal with reduced noise. These properties allow for the difference rewards to significantly boost learning performance in a MAS.

The challenge for deriving the difference rewards signal is obviously how to calculate the second term of the equation $G(z_{-i})$ which is called the \textit{counterfactual}. In particular domains, the counterfactual $G(z_{-i})$ is possible to be directly calculated \cite{tumer2009,devlin2014}. Alternatively, in cases where this is not possible, it has been demonstrated that difference rewards can be estimated \cite{tumer2007,proper2013}.

Difference rewards have been successfully applied in different domains such as congestion problems \cite{tumer2007,tumer2009,proper2013,devlin2014}, network security \cite{malialis2015cs} and distributed sensor networks \cite{colby2013}.

\section{Congestion Problems as Multiagent Learning Problems}\label{sec:formulation}
This section provides a formal definition of congestion problems as multiagent learning problems. A congestion problem occurs when multiple entities are competing for a limited amount of available resources. Different congestion problems (traffic congestion, air traffic management, Beach problem, El Farol Bar problem) have been formulated in the past as multiagent learning problems \cite{tumer2007,tumer2009,proper2013,devlin2014}.

Formally, there exists a set $S$ of $n$ resources $S =$\\$\{s_{1}, s_{2}, ..., s_{n}\}$. Each resource $s$ is a 3-tuple $s = <w_{s},\psi_{s},x_{s,t}>$ where $w_{s} \geq 0$ is the weighting or importance of resource $s$, $\psi_{s} > 0$ is the capacity of resource $s$, and $x_{s,t} \geq 0$ is the consumption of resource $s$ at time $t$. A resource $s$ at time $t$ is said to be congested if its consumption is higher than its capacity i.e. $x_{s,t} > \psi_{s}$.

The local reward ($L$) of an entity located in resource $s$ at time $t$ depends on the resource's weighting, capacity and consumption as shown in Equation~\ref{eq:L}:
\begin{eqnarray}\label{eq:L}
L(s,t) = f(w_{s}, \psi_{s}, x_{s,t})
\end{eqnarray}

The global reward ($G$) is defined as the summation over all local rewards and is given by Equation~\ref{eq:G}.
\begin{eqnarray}\label{eq:G}
G(t) = \sum_{s \in S} L(s,t)
\end{eqnarray}

Recall from Section~\ref{sec:D} that the difference rewards signal is calculated by subtracting the counterfactual from the global reward. Therefore, the difference rewards ($D$) for agent $i$ located in section $s$ at time $t$ can be calculated using Equation~\ref{eq:D}.
\begin{eqnarray}\label{eq:D}
D_i(t) & = & G(t) - G_{-i}(t) \notag\\
	   & = & L(s,t) - L_{-i}(s,t)
\end{eqnarray}

The straightforward approach is to consider the entities which compete for the resources as the learning agents. For instance, in a traffic congestion domain \cite{tumer2009} where the entities are cars which compete for traffic lanes, we can install a learning agent on each car. In such domains, there is no need to estimate the counterfactual as it can simply be calculated by decreasing the resource's consumption or attendance by 1 as shown in Equation~\ref{eq:D2}
\begin{eqnarray}\label{eq:D2}
D_i(t) & = & L(s,t) - L_{-i}(s,t) \notag\\
	   & = & f(w_{s}, \psi_{s}, x_{s,t}) - f(w_{s}, \psi_{s}, x_{s,t}-1)
\end{eqnarray}

Alternatively, there exist congestion problems where learning agents are not installed on the entities competing for the resources and in such cases the counterfactual needs to be estimated \cite{tumer2007}. Although estimating $D_i$ has been successfully demonstrated in a number of applications this topic is under ongoing research \cite{proper2013}.

\section{Resource Abstraction}\label{sec:abstract}
The idea behind our approach is to allocate the available resources into groups called abstract groups which provide a more informative signal to the learning agents.

Formally, there exists a set $B$ of $p$ abstract groups: $B = \{b_{1}, b_{2}, ..., b_{p}\}$. Each abstract group $b$ is a 4-tuple:\\$b=<M_{b},W_{b},\Psi_{b},X_{b,t}>$.

The set $M_{b} = \{m_{1}, m_{2}, ..., m_{k}\}$ is the member set of abstract group $b$ that consists of $k$ resources. A resource $m$ can belong only to one abstract group, that is, $m \in M_i$ and $m \in M_j$ only if $i = j$. The number of abstract groups (and their members) depends on the application domain.

The weight $W_{b}$ of abstract group $b$ is defined as the average weight of its members as shown in Equation~\ref{eq:W}.
\begin{eqnarray}\label{eq:W}
W_{b} = \frac{1}{k} \sum_{m \in M_{b}} w_{m}
\end{eqnarray}

The capacity $\Psi_{b}$ of abstract group $b$ is the total capacity of its members. Similarly, the consumption $X_{b,t}$ of abstract group $b$ at time $t$  is the total consumption of its members. These are calculated using Equations~\ref{eq:Psi} and~\ref{eq:X}.
\begin{eqnarray}\label{eq:Psi}
\Psi_{b} = \sum_{m \in M_{b}} \psi_{m}
\end{eqnarray}

\begin{eqnarray}\label{eq:X}
X_{b,t} = \sum_{m \in M_{b}} x_{m,t}
\end{eqnarray}

As discussed, the straightforward approach is to consider the entities which compete for the resources as the learning agents. However, depending on the application, this may not necessarily be the case, and our proposed approach does not impose any limitations on the agent selection.

Let us now introduce the proposed reward functions. We start by extending the local reward $L(s,t)$, and define the signal $H(b,t)$ for an abstract group $b$ at time $t$. It is calculated by taking into account the abstract group's weight, capacity and consumption as shown in Equation~\ref{eq:H}. It is important to note that the function $f(\cdot)$ is the same which is used to calculate the local reward $L$ (see Equation~\ref{eq:L}).
\begin{eqnarray}\label{eq:H}
H(b,t) = - f(W_{b}, \Psi_{b}, X_{b,t})
\end{eqnarray}

The \textbf{abstract reward ($A$)} is defined as follows. When resource $s$ which belongs to abstract group $b$ is not congested, we provide a reward from $L(s,t)$. Alternatively, we provide a reward from the abstraction configuration i.e. from $H(b,t)$. This is shown in Equation~\ref{eq:A}.
\begin{eqnarray}\label{eq:A}
 A(b,s,t) =
  \begin{cases}
   L(s,t) & \text{if } x_{s,t} \leq \psi_{s} \\
   H(b,t) & \text{if } x_{s,t} > \psi_{s}
  \end{cases}
\end{eqnarray}
\noindent where $s \in M_{b}$.

The intuition behind this reward function is to facilitate the coordination among the learning agents by providing them with a more informative signal. Assuming a ``good'' abstraction configuration (as mentioned, this is application domain-specific), the abstract reward ($A$) will provide a bigger (than $L$) punishment to the agents to encourage them to keep resources decongested. Furthermore, as we will later demonstrate, in domains where the number of entities competing for the resources is greater than the number of available resources, the desired solution is to congest only a minimal amount of them (e.g. only one) and leave the rest decongested. The approach will encourage some agents to consume a congested resource. In such cases the agents already receive a punishment, but they will be provided with a bigger punishment if the agents attempt to switch to a decongested abstract group and congest it.\\\\\\

\section{Beach Problem Domain (BPD)}\label{sec:BPD}

\subsection{Domain Description}
The Beach Problem Domain (BPD) \cite{devlin2014} is an abstract congestion problem that relates to many real-life congestion problems and allows us to perform a thorough analysis of our proposed approach. In this domain, each tourist (learning agent) gets to choose the beach section it will visit. At each time step each agent knows which beach section it is currently on and must choose to either stay still or move to an adjacent beach section (i.e. left or right).

When all the agents have performed their actions, they receive a reward. The highest reward for a beach section is received when the attendance of agents is equal to the capacity of the beach section. If a beach section gets congested or overcrowded it is undesirable. Beach sections with low attendance are also undesirable. This constitutes a congestion problem when the number of agents is much greater than the total capacity of the beach sections.

In this abstract congestion problem it is assumed that all beach sections have an equal weighting of 1 and they all have the same capacity $\psi$. The local reward function ($L$) is given by Equation~\ref{eq:BPD-L}:
\begin{eqnarray}\label{eq:BPD-L}
L(s,t) = x_{s,t} \mathrm{e}^{\frac{-x_{s,t}}{\psi}}
\end{eqnarray}

\noindent where $s$ is the beach section and $x_{s,t}$ is the attendance of beach section $s$ at time step $t$.

The global reward ($G$) is a summation over all local rewards and is given by Equation~\ref{eq:BPD-G}.
\begin{eqnarray}\label{eq:BPD-G}
G(t) = \sum_{s \in S} x_{s,t} \mathrm{e}^{\frac{-x_{s,t}}{\psi}}
\end{eqnarray}

The difference rewards ($D$) signal is calculated by applying Equation~\ref{eq:D2} to Equation~\ref{eq:BPD-L}. This results in Equation~\ref{eq:BPD-D}.
\begin{eqnarray}\label{eq:BPD-D}
D_i(t) = x_{s,t} \mathrm{e}^{\frac{-x_{s,t}}{\psi}} - (x_{s,t}-1) \mathrm{e}^{\frac{-(x_{s,t}-1)}{\psi}}
\end{eqnarray}

Similarly to the local reward function, we define the corresponding reward function for an abstraction configuration $B$ (defined as described in Section~\ref{sec:abstract}). Therefore, the application of Equation~\ref{eq:H} to Equation~\ref{eq:BPD-L} results in Equation~\ref{eq:BPD-H}:
\begin{eqnarray}\label{eq:BPD-H}
H(b,t) = - X_{b,t} \mathrm{e}^{\frac{-X_{b,t}}{\Psi_{b}}}
\end{eqnarray}

\noindent where $b$ is the abstract group, $X_{b,t}$ is the attendance at time step $t$, and $\Psi_{b}$ is the capacity of abstract group $b$.

Lastly, the abstract reward ($A$) is calculated by applying Equation~\ref{eq:A} and is shown in Equation~\ref{eq:BPD-A}.
\begin{eqnarray}\label{eq:BPD-A}
 A(b,s,t) =
  \begin{cases}
   x_{s,t} \mathrm{e}^{\frac{-x_{s,t}}{\psi}} & \text{if } x_{s,t} \leq \psi \\
   - X_{b,t} \mathrm{e}^{\frac{-X_{b,t}}{\Psi_{b}}} & \text{if } x_{s,t} > \psi
  \end{cases}
\end{eqnarray}

\noindent where $s \in M_{b}$.

\begin{algorithm}[t]             
\caption{Beach Problem with Resource Abstraction}   
\label{alg:BPD}                  
\begin{algorithmic}[1]           
\STATE{define abstraction configuration $B$}
\STATE{initialise Q-values: $\forall s,a | Q(s,a) = -1$}
\STATE{set agents to initial locations}
\FOR{$episode = 1:num\_episodes$}
	\FOR{$timestep = 1:num\_timesteps$}
		\FOR{$agent = 1:num\_agents$}
			\STATE{perceive current beach section $s \in S$}
			\STATE{choose action $a=\{-1,0,+1\}$ using $\epsilon$-greedy}
			\STATE{move to section $s^{\prime}=\{s-1,0,s+1\}$}
			\IF{$s^{\prime} \notin S$}
			\STATE{move to nearest section $s^{\prime} \in S$}
			\ENDIF
		\ENDFOR
		
		\FOR{$section = 1:num\_sections$}
			\STATE{calculate $A$ reward (Equation~\ref{eq:BPD-A})}
		\ENDFOR
		
		\FOR{$agent = 1:num\_agents$}
			\STATE{update Q(s,a) using A (Equation~\ref{eq:qlearning})}
		\ENDFOR
		
		\STATE{reduce $\alpha$ using $alpha\_decay\_rate$}
		\STATE{reduce $\epsilon$ using $epsilon\_decay\_rate$}
	\ENDFOR
	\STATE{reset agents to initial locations}
\ENDFOR
\end{algorithmic}
\end{algorithm}

Algorithm~\ref{alg:BPD} presents what has been described thus far. Before describing the experimental work, we should note that the best solution for this problem is to overcrowd one (any) beach section, and leave $\psi$ number of agents in the rest of the sections. This will give the highest possible system performance or global reward.\\

\subsection{Experimental Setup}\label{sec:setup}
Our experimental setup is as follows. The learning rate is set to $\alpha = 0.1$ and $alpha\_decay\_rate=0.9999$. The exploration parameter is set to $\epsilon=0.05$ and $epsilon\_decay\_rate=0.9999$. The discount factor is set to $\gamma=1.0$ and the number of episodes is set to $num\_episodes=10000$. Initially, the agents are uniformly distributed (i.e. it is assumed that there exist hotels for the tourists in each beach section). The rest of the parameters are given later for each experiment. In all experiments, we plot the system performance/global reward $G$ at the last time step of each episode. The values are averaged over 30 statistical runs and error bars showing the standard error around the mean are plotted. In some plots the error bars are very small and hardly visible, but they are present on all plots.

\subsection{Experimental Results}
We have set $num\_agents=100$, $num\_sections=6$, capacity $\psi = 6$ and $num\_timesteps = 5$. The highest social welfare or global reward is 11.04 and occurs when one (any) of the six sections is overcrowded with 70 agents, while each of the remaining five sections have six agents; this is shown with a black dashed line in all figures.

\begin{figure}[t]
\centering
\includegraphics[scale=0.4]{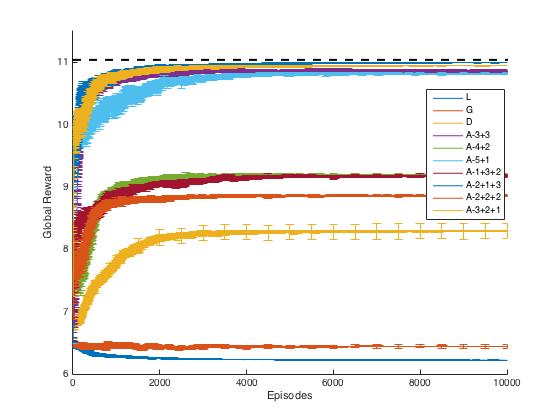}
\caption{Abstraction configurations}\label{fig:1}
\end{figure}

The first experimental study investigates how different abstraction ($A$) configurations affect the performance, and how they compare to the approaches which use the local ($L$), global ($G$) and difference ($D$) rewards. Figure~\ref{fig:1} demonstrates this with abstraction configurations of two and three abstract groups. For example, the plot with label ``A-4+2'' means that the first four beach sections are grouped into one abstract group, while the remaining two beach sections are grouped into another abstract group. There are four important outcomes from this study:
\begin{itemize}
\item Abstraction selection does affect the system performance.
\item However, all the abstractions outperform the system that uses the difference ($D$) rewards (third plot from the bottom).
\item The abstraction ``A-2+1+3'' achieves the highest performance while three other achieve a near-highest performance. The system using $D$ never achieves the highest social welfare.
\item All the abstractions outperform the systems that use the local ($L$) and global ($G$) rewards (bottom two plots).
\end{itemize}

The second experimental study examines how the episode length i.e. number of time steps per episode affects the system performance. Recall that at the start of each new episode, that is, after the last time step of the previous episode (Algorithm~\ref{alg:BPD}/Line $23$) the agents return to their initial location. It is important to note that changing the length of the episode produces a different and independent version of the problem.

\begin{figure}[t]
\centering
\includegraphics[scale=0.4]{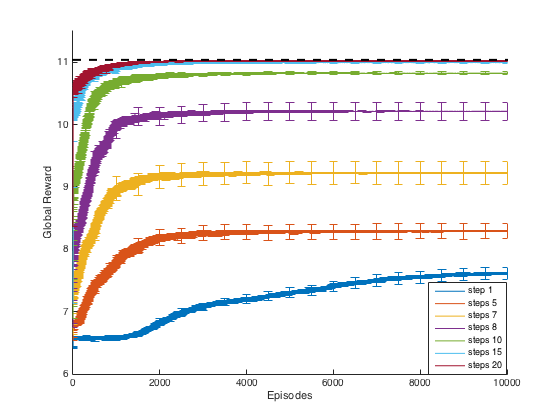}
\caption{Impact of episode length on D}\label{fig:2}
\end{figure}

In Figure~\ref{fig:2} we apply $D$ in different versions of the BPD, where we vary the time (episode length) agents are allowed to settle to a beach section within an episode. Note that with five or more time steps per episode, an agent located at the far left (or right) beach section can, in principle, move to the far right (or left) section at the end of an episode. However, $D$ achieves the highest performance only in BPD versions with episode length of 15 time steps or more.

\begin{figure}[t]
\centering
\includegraphics[scale=0.4]{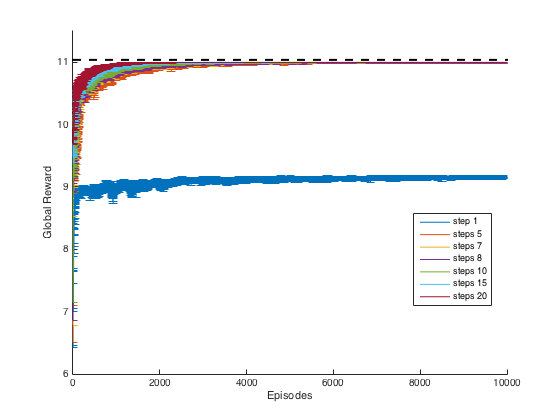}
\caption{Impact of episode length on A}\label{fig:3}
\end{figure}

We now run the same experiment for the proposed Resource Abstraction approach. Based on previous findings (Figure~\ref{fig:1}) we proceed with abstraction ``A-2+1+3''. Figure~\ref{fig:3} depicts the performance of $A$ in different BPD versions. It is observed that the highest social welfare is achieved in all BPD versions except when a single time step is used. Also, $A$ learns faster and achieves a significantly higher final performance than $D$ in all BPD versions except from versions with episode length of 15 steps or more, where both have a similar performance. These are important results for potential real-world applications as they show that $A$ is more robust to changes in the problem domain (i.e. episode length) than the current state-of-the-art approach $D$.

The third experimental study of BPD aims at investigating the scalability of the proposed approach. The setup for this study involves $num\_agents=1000$, $num\_sections=20$, capacity $\psi = 18$ and $num\_timesteps = 5$. The highest social welfare is 125.82 and is shown with a black dashed line in all figures.

\begin{figure}[t]
\centering
\includegraphics[scale=0.4]{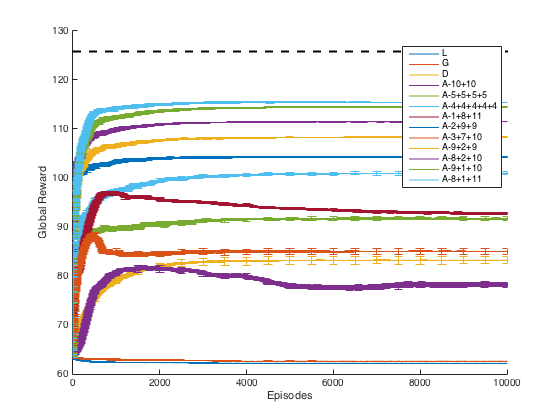}
\caption{BPD with 1000 agents, 5 time steps}\label{fig:4}
\end{figure}

Like before, we have investigated a range of abstraction configurations and many of them are shown in Figure~\ref{fig:4}. We observe the following. All the abstractions except one (third plot from the bottom) outperform the system that uses the difference ($D$) rewards (fourth plot from the bottom). All the abstractions outperform the systems that use the local ($L$) and global ($G$) rewards (bottom two plots). Note that no approach manages to achieve the best performance in just five time steps per episode. However, taking into consideration the scale of the scenario (1000 agents, 20 sections), some of the abstractions (see upper plots in Figure~\ref{fig:4}) perform extremely well.

\begin{figure}[t]
\centering
\includegraphics[scale=0.4]{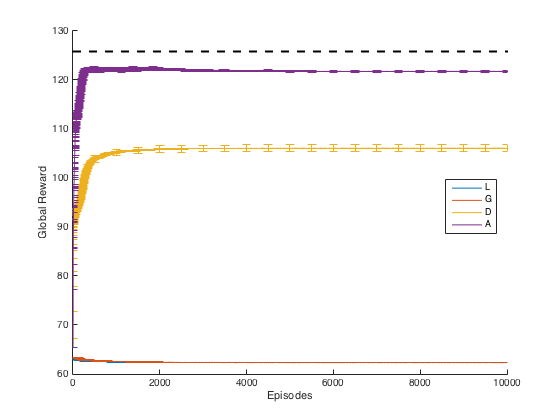}
\caption{BPD with 1000 agents, 10 time steps}\label{fig:5}
\end{figure}

\begin{figure}[t]
\centering
\includegraphics[scale=0.4]{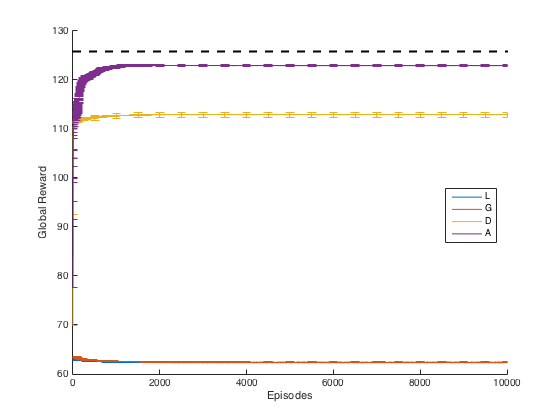}
\caption{BPD with 1000 agents, 20 time steps}\label{fig:6}
\end{figure}

We proceed with abstraction ``A-8+1+11'' and repeat the same experiment with ten and 20 time steps per episode; results are shown in Figures~\ref{fig:5} and~\ref{fig:6} respectively. In both figures, the system using $A$ achieves a near-highest performance. The system using $D$ performs orders of magnitude better than the local ($L$) and global ($G$) rewards, but still it is significantly outperformed by the proposed $A$ approach.\\

\section{Traffic Lane Domain (TLD)}\label{sec:TLD}

\subsection{Domain Description}
Despite the BPD being an abstract congestion problem, it is very important as it relates to many real-world congestion problems. One such complex problem is the Traffic Lane Domain (TLD) \cite{tumer2009} which is defined as follows.

In this domain each car (learning agent) gets to choose which traffic lane to follow. Each agent knows at each time step which lane it currently follows. At each time step, each agent can either stay in the same lane, or move to an adjacent lane (i.e. left or right). Once all the agents have executed their actions, they receive a reward.

To simulate real-life characteristics there are three major differences with the BPD: \textit{1.} traffic lanes have a different weighting (represents the different preference of drivers) \textit{2.} lanes have different capacities (represents the fact that lanes are of different sizes, or that there exist restrictions due to construction works, tolls, carpools etc.) and \textit{3.} it has a different reward function as it doesn't matter how many cars are on a lane as long as the lane is not congested (as opposed to a beach section which is desirable to have more people as long as it is not overcrowded).

The local reward ($L$) function is given in Equation~\ref{eq:TLD-L}:
\begin{eqnarray}\label{eq:TLD-L}
L(s,t) =
\begin{cases}
w_{s} \mathrm{e}^{-1} & \text{if } x_{s,t} \leq \psi_{s} \\
w_{s} \mathrm{e}^{\frac{-x_{s,t}}{\psi_{s}}} & \text{if } x_{s,t} > \psi_{s}
\end{cases}
\end{eqnarray}

\noindent where $s$ is the traffic lane, $w_s$ is the weighting, $x_{s,t}$ is the attendance at time step $t$ and $\psi_s$ is the capacity.

The global ($G$) and difference ($D$) rewards are calculated using Equations~\ref{eq:G} and~\ref{eq:D2} respectively.

The application of Equation~\ref{eq:H} to Equation~\ref{eq:TLD-L} is shown in Equation~\ref{eq:TLD-H}:
\begin{eqnarray}\label{eq:TLD-H}
H(b,t) =
\begin{cases}
- W_{b} \mathrm{e}^{-1} & \text{if } X_{b,t} \leq \Psi_{b} \\
- W_{b} \mathrm{e}^{\frac{-X_{b,t}}{\Psi_{b}}} & \text{if } X_{b,t} > \Psi_{b}
\end{cases}
\end{eqnarray}

\noindent where $b$ is the abstract group, $W_{b}$ is the weighting, $X_{b,t}$ is the attendance at time step $t$, and $\Psi_b$ is the capacity. Lastly, the abstract reward ($A$) is calculated using Equation~\ref{eq:A}.

\subsection{Experimental Setup}
We have set $\alpha = 0.1$, $alpha\_decay\_rate=0.9999$, $\epsilon=0.05$, $epsilon\_decay\_rate=0.9999$, $\gamma=1.0$ and $num\_episodes=10000$. Initially, the agents are uniformly distributed. The rest of the parameters are given below. In all experiments, we plot the system performance/global reward $G$ at the last time step of each episode (except where otherwise is stated). The values are averaged over 30 statistical runs and error bars showing the standard error around the mean are plotted. In some plots the error bars are very small and hardly visible, but they are present on all plots.

\subsection{Experimental Results}
For the first experimental study in the TLD we adopt the same traffic scenario as in \cite{tumer2009}. In this scenario, there are 500 cars or learning agents and 9 traffic lanes each with capacity 167, 83, 33, 17, 9, 17, 33, 83, 167 (i.e. a total of 609). The weightings for each lane are 1, 5, 10, 1, 5, 10, 1, 5, 10. The highest possible system performance or global reward is 17.66 and is achieved when all the lanes are decongested; this is shown with a black dashed line in all figures.
 
The fact that there are different weightings for each resource or traffic lane gives us the opportunity to try non-contiguous abstractions i.e. abstract groups with non-adjacent members. The obvious configuration is to have three abstract groups each with three member lanes. The first, second and third abstract groups include lanes with weighting 1, 5, and 10 respectively. We will refer to this abstraction configuration as ``A (non-contig)''. Moreover, like in the BPD, we have experimented with many abstraction configurations and found out that the best performance is obtained with ``A-1+8'', which we will refer to as ``A (contig)''.

\begin{figure}[t]
\centering
\includegraphics[scale=0.4]{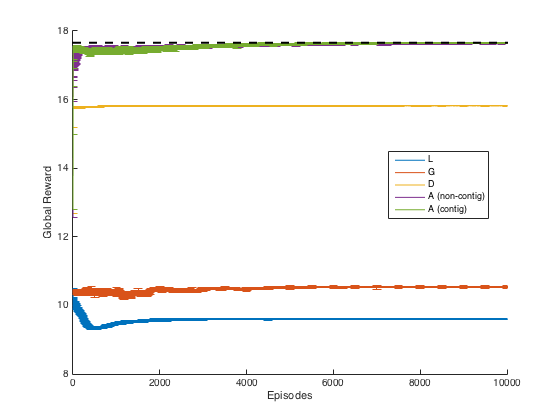}
\caption{TLD with 500 cars, 5 time steps}\label{fig:7}
\end{figure}

\begin{figure}[t]
\centering
\includegraphics[scale=0.4]{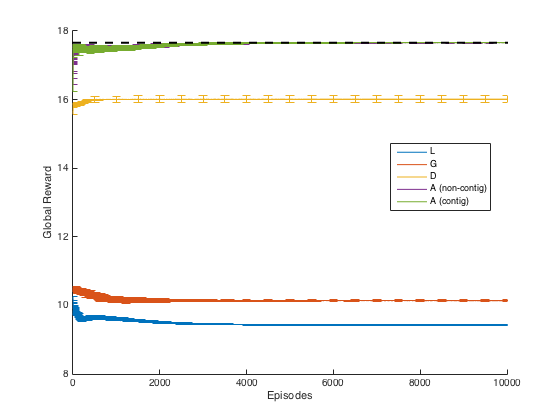}
\caption{TLD with 500 cars, 10 time steps}\label{fig:8}
\end{figure}

Figure~\ref{fig:7} shows the performance for all approaches for five time steps per episode. The system using abstract rewards ($A$) achieves the highest performance while both $A$ and $D$ completely outperform $L$ and $G$. The same experiment with ten time steps per episode is repeated in Figure~\ref{fig:8} where very similar results are obtained. Note that with ten time steps, an agent located at the far left (or right) lane has, in principle, adequate time to reach the far right (or left) end. Despite the increase in the number of time steps, the system performance using $D$ is only slightly improved.

An important question is how will the system perform if the lane weightings and capacities change in due course e.g. because of an accident? Consequently, a crucial question that certainly has deployment implications is will a new abstraction configuration be needed each time the lane weightings and capacities change?

The second experimental study in the TLD provides answers to these questions. To simulate such a behaviour we assume that two accidents occurred in lanes 3 and 9 at episode 2000. As a result, their capacity is reduced to half. Also, each of the two lanes has swapped weighting (preference) with one of its adjacent lanes. Therefore, the new capacities are 167, 83, 17, 17, 9, 17, 33, 83, 83 (i.e. a total of 509) and the new weightings are 1, 10, 5, 1, 5, 10, 1, 10, 5. We keep the same number of cars i.e. 500 and at the time of the accidents the agents' exploration parameter is reset to the initial. The number of time steps per episode is five.

\begin{figure}[t]
\centering
\includegraphics[scale=0.4]{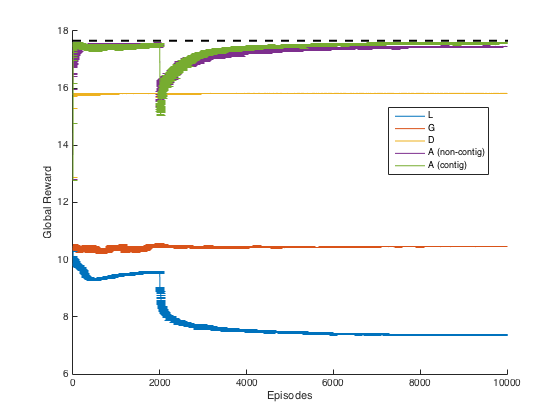}
\caption{TLD with accidents}\label{fig:9}
\end{figure}

Figure~\ref{fig:9} shows the results for the scenario with the accidents. The system using $L$ is affected the most and does not manage to recover i.e. to reach the performance it had achieved before the occurrence of the accidents. Interestingly the system using $G$ is not affected at all, but its performance is poor. The system using $A$ is affected but manages to completely recover and achieve the highest possible welfare as it used to be the case without the occurrence of the accidents. The system using $D$ is also affected but only slightly (it is not visible in the graph as it is out-shaded by the decline of the $A$ plots), however, the only system that achieves the best performance is the one using $A$. The outcome of this study suggests that a system using the proposed approach is robust and does not need to refine the abstraction configuration if something is altered in due course such as the traffic lane capacities and weightings.

It has been assumed so far that all learning agents or cars are participating in the learning scheme. The third experimental study for the TLD examines the behaviour of the approaches when this assumption is violated. In practise, $100\%$ compliance by drivers is unlikely to be achieved \cite{tumer2009}. This is because some drivers may not be convinced to participate or even if they are all willing to participate, some of them may come across different problems such as information and sensing limitations.

\begin{figure}[t]
\centering
\includegraphics[scale=0.4]{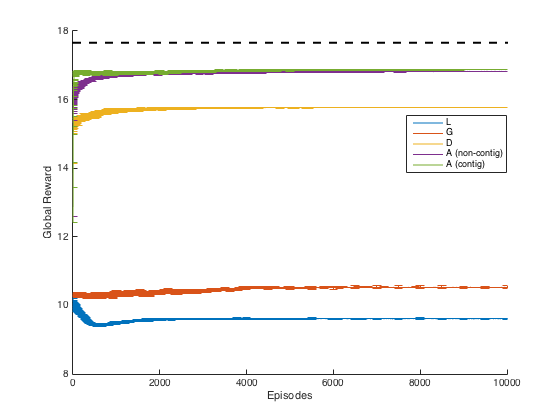}
\caption{TLD with 25\% non-compliant drivers}\label{fig:10}
\end{figure}

Figure~\ref{fig:10} depicts the system performance ($num\_agents=500$, $num\_timesteps=5$) in the presence of $25\%$ non-compliant drivers. Non-compliant drivers are simulated as agents sticking to their initial choice/location. It is observed that the performance of $L$ and $G$ is not affected at all, but it is very poor. The performance of $D$ and $A$ is not affected much and, as a result, the system using $A$ still outperforms the system that uses $D$. The results suggest that the proposed system works well under non-compliant drivers.

The fourth experimental study further investigates the potential deployment of such a learning scheme. In practise, it is expected that such a system will be rolled out in phases \cite{tumer2009}. For instance, consider a scenario of four delivery phases where the participation is 25\%, 50\%, 75\% and 100\% of the drivers. To simulate this situation we vary the rate of non-compliant drivers, for example, the first phase with 25\% participation corresponds to 75\% non-compliant drivers.

\begin{figure}[t]
\centering
\includegraphics[scale=0.4]{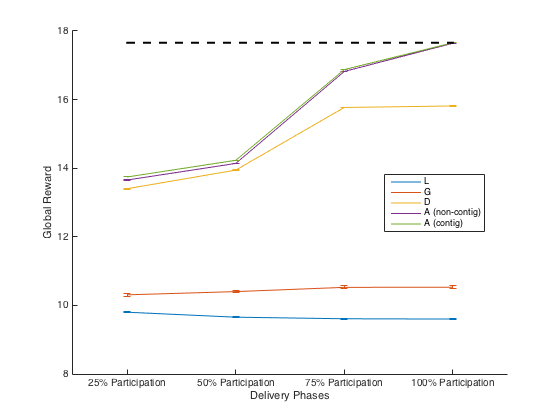}
\caption{TLD with 4 delivery phases}\label{fig:13}
\end{figure}

Figure~\ref{fig:13} shows the converged performance ($num\_agents=500$, $num\_timesteps=5$) for the scenario with four delivery phases where the superiority of the proposed approach is clearly demonstrated, suggesting that the proposed system can be introduced to the public in different stages.

\begin{figure}[t]
\centering
\includegraphics[scale=0.4]{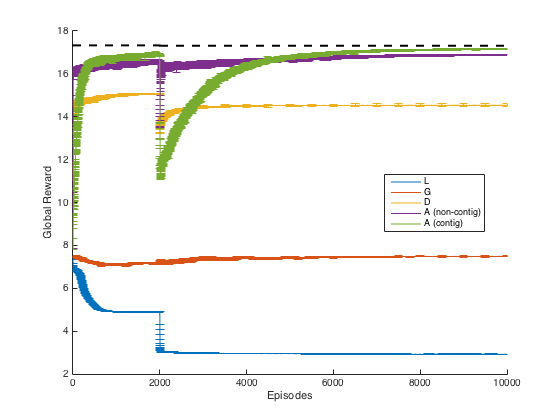}
\caption{TLD (1000 cars) with accidents}\label{fig:15}
\end{figure}

In the last study we repeat the experiment with 1000 cars. The highest performance is 17.33 and occurs when the first lane (capacity 167, weighting 1) is overcrowded with 558 agents while the remaining eight lines have full capacity. We repeat the case under accident conditions (as in Figure~\ref{fig:9}), where the highest performance is 17.31 and occurs when the first lane is overcrowded with 658 agents. Figure~\ref{fig:15} shows the results ($num\_timesteps=5$) where the superiority and the ability of the proposed approaches to recover is shown.

\section{Discussion and Conclusion}\label{sec:conclusion}
We have introduced Resource Abstraction for multiagent coordination problems. Our approach allocates available resources into abstract groups to provide a more informative signal to learning agents, significantly improving both learning speed and social welfare.

Experimental work was conducted on two benchmark domains, an abstract congestion problem and a realistic traffic congestion problem. The state-of-the-art solution to solve multiagent congestion problems is a form of reward shaping called difference rewards. We have shown that our novel approach of resource abstraction requires significantly fewer time steps per learning episode to achieve the highest social welfare. This is important for real-world applications where the availability of learning time is limited.

Furthermore, it has been empirically demonstrated in both experimental domains that the proposed approach not only performs better than the two baseline and the difference rewards approaches, but it achieves the highest possible or near-highest system performance even in experiments involving 1000 learning agents. Scalability is another strong aspect of the proposed approach. It has been shown that the proposed approach is robust against changes (e.g. car accidents in the traffic congestion domain) and capable of operating without 100\% agent compliance.

It has been shown that when the number of learning steps is limited, then in almost every case Resource Abstraction outperforms the state-of-the-art irrespective of the abstraction grouping (Figures~\ref{fig:1} and~\ref{fig:4}). When the number of learning steps is not limited, we suggest some guidelines that help us define the abstraction grouping. Some experimentation is still necessary but the following heuristics simplify this process: \textit{a.} three abstract groups typically suffice as this allows more resources to be included in an abstract group, hence, a larger penalty can be provided if the group becomes congested \textit{b.} group resources with similar characteristics e.g. the same weightings; the effectiveness of this heuristic is shown by \textit{A (non-contig)} in Figures~\ref{fig:7} -~\ref{fig:15}.

Moreover, it has been shown that if something unexpected occurs (e.g. accidents in the traffic lane domain) it is highly likely that there will be no need to refine the abstraction grouping, as, given some additional exploration time, the system can recover and still achieve the highest performance.

%
%
\bibliographystyle{abbrv}
\bibliography{aamas2016}  

\begin{thebibliography}{10}

\bibitem{claus1998}
C.~Claus and C.~Boutilier.
\newblock The dynamics of reinforcement learning in cooperative multiagent
  systems.
\newblock In {\em Proceedings of the AAAI Conference on Artificial
  Intelligence}, 1998.

\bibitem{colby2013}
M.~Colby and K.~Tumer.
\newblock Multiagent reinforcement learning in a distributed sensor network
  with indirect feedback.
\newblock In {\em Proceedings of the 2013 international conference on
  Autonomous agents and multi-agent systems}, pages 941--948, 2013.

\bibitem{devlin2014}
S.~Devlin, L.~Yliniemi, K.~Tumer, and D.~Kudenko.
\newblock Potential-based difference rewards for multiagent reinforcement
  learning.
\newblock In {\em Proceedings of the 13th International Conference on
  Autonomous Agents and Multiagent Systems}, 2014.

\bibitem{malialis2015cs}
K.~Malialis, S.~Devlin, and D.~Kudenko.
\newblock Distributed reinforcement learning for adaptive and robust network
  intrusion response.
\newblock {\em Connection Science}, 27(3):234--252, 2015.

\bibitem{mataric1998}
M.~J. Matari{\'c}.
\newblock Using communication to reduce locality in distributed multi-agent
  learning.
\newblock {\em Journal of Experimental and Theoretical Artificial Intelligence,
  special issue on Learning in DAI Systems, Gerhard Weiss, ed.},
  10(3):357--369, 1998.

\bibitem{proper2013}
S.~Proper and K.~Tumer.
\newblock Multiagent learning with a noisy global reward signal.
\newblock In {\em Proceedings of the AAAI Conference on Artificial
  Intelligence}, 2013.

\vfill\eject

\bibitem{stone2007}
P.~Stone.
\newblock Learning and multiagent reasoning for autonomous agents.
\newblock In {\em Proceedings of the International Joint Conference on
  Artificial Intelligence (IJCAI)}, pages 12--30, 2007.

\bibitem{stone2000}
P.~Stone and M.~Veloso.
\newblock Multiagent systems: A survey from a machine learning perspective.
\newblock {\em Autonomous Robots}, 8(3):345--383, 2000.

\bibitem{sutton1998}
R.~S. Sutton and A.~G. Barto.
\newblock {\em Introduction to Reinforcement Learning}.
\newblock MIT Press Cambridge, MA, USA, 1998.

\bibitem{tumer2007}
K.~Tumer and A.~Agogino.
\newblock Distributed agent-based air traffic flow management.
\newblock In {\em Proceedings of the 6th International Joint Conference on
  Autonomous Agents and Multiagent Systems}, 2007.

\bibitem{tumer2009}
K.~Tumer, Z.~Welch, and A.~K. Agogino.
\newblock Traffic congestion management as a learning agent coordination
  problem.
\newblock In A.~Bazzan and F.~Kluegl, editors, {\em Multiagent Architectures
  for Traffic and Transportation Engineering}, pages 261--279. Lecture notes in
  AI, Springer, 2009.

\bibitem{watkins1992}
C.~J. Watkins and P.~Dayan.
\newblock Q-learning.
\newblock {\em Machine learning}, 8(3-4):279--292, 1992.

\bibitem{wolpert2000}
K.~R.~W. Wolpert, David~H. and K.~Tumer.
\newblock Collective intelligence for control of distributed dynamical systems.
\newblock {\em Europhysics Letters}, 49(6), 2000.

\end{thebibliography}
%
\end{document}